# Quantum Well Electronic States in Spatially Decoupled 2D Pb Nanoislands on Nb-doped SrTiO$_3$(001)

Guan-Yu Chen[1], Chia-Hsiu Hsu[2], Bo-You Liu[1], Li-Wei Chang[1], Deng-Sung Lin[1,3], Feng-Chuan Chuang[2,1,4] and Pin-Jui Hsu[1,3,*]

[1]Department of Physics, National Tsing Hua University, Hsinchu 30013, Taiwan

[2]Department of Physics, National Sun Yat-sen University, Kaohsiung 804, Taiwan

[3]Center for Quantum Technology, National Tsing Hua University, Hsinchu 30013, Taiwan

[4]Physics Division, National Center for Theoretical Sciences, Hsinchu, Taiwan

[*]Corresponding authors: pinjuihsu@phys.nthu.edu.tw (P. J. Hsu)

**Abstract**

Two-dimensional (2D) Pb nanoisland has established an ideal platform for studying the quantum size effects on growth mechanism, electronic structures as well as high-temperature superconductivity. Here, we investigate the growth and quantum well electronic states of the 2D Pb nanoisland on Nb-doped SrTiO$_3$(001) by scanning tunneling microscopy and spectroscopy. In contrast to Pb/Si(111), Pb/Cu(111) and Pb/Ag(111), there is no wetting layer of Pb formed on Nb-doped SrTiO$_3$(001) surface, resulting in isolated Pb nanoislands to initiate the growth from building blocks of an apparent height with 4 atomic layers. According to the thickness-dependent quantum well states resolved in both occupied and unoccupied energy regions, the constant group velocity $v_g$ = 1.804 × 10$^6$ m/s and Fermi wavevector $k_F$ = 1.575 Å$^{-1}$, have been extracted from a linear fit of the Pb(111) band dipersion along the $\Gamma - L$ direction. In addition, the energy-dependent scattering phase shift $\varphi(E)$ obtained by means of phase accumulation model shows a metallic-like scattering interface as compared to Pb/Ag(111). These spatially decoupled 2D Pb nanoislands thus realize an opportunity to explore the intrinic quantum confinement phenomena in nanoscale superconductors on the doped titanium-oxide-type substrate.

**Introduction:**

Since the discovery of high-temperature superconductivity in one-unit-cell FeSe film grown on Nb-doped SrTiO$_3$(001) (Nb-STO)[1], i.e., superconducting transition temperature $T_C$ dramatically increases up to 109 K, the enhancement of $T_C$ by interface effects has been reported on various Fe-based superconducting ultrathin films prepared on pervoskite-type titanium oxide substrates[2-5]. Besides the unconventional Fe-based superconductors, the elementary β-Sn islands self-assembled on the Nb-STO(001) substrate also exhibit an improved $T_C$ up to 8.2 K as compared to the bulk $T_C$ about 3.7 K, which has been attributed to the interfacial charge transfer as well as enhanced electron-phonon coupling at the nanometer-sized β-Sn islands[6].

Recently, the 2D Pb nanoislands have been fabricated with three different types of apparent heights as well as area sizes on Nb-STO(001) substrate, and they exhibit the superconducting $T_c$ about 7.0 K close to the value of bulk Pb material[7]. Although nearly the same $T_c$ with bulk Pb, these 2D Pb nanoislands grown on Nb-STO(001) show an enlarged superconducting gap $\Delta_S$, giving rise to unusual strong coopuling based on an enhanced BCS ratio ($2\Delta_S/k_B T_C$) ~ 6.22. In contrast to $T_c$ about 7.0 K, Sun et al. reported a rather high $T_c$ up to 47 K on these 2D Pb nanoislands[8], but their results might be interfered with the induced Coulomb gap in the Pb islands particularly with a reduced dimensionality[7].

In addition to pseudogap feature induced by Coulomb interaction aforementioned[7], quantum size effect typically plays a dominant role in quantum confinement of electronic states of the 2D Pb nanoislands grown Nb-STO(001) or Pb films grown on different substrates, leading to discrete energy levels so called quantum well states (QWSs)[9-11]. The formation of these QWSs has been well known for substantial influences on physical and chemical properties of nanostructures, for examples, electrical resistivity[12], superconducting transition temperature[13-15], thermal stability[16,17] and chemical reactivity[18] etc. Given that discrete QWSs developed in Pb nanostructures are typically confined by a finite barrier height from two boundaries, i.e., vacuum and substrate, the electron waves could have a chance to penetrate into these two regions. While such electron wave penetration is inevitable to occur, the phase shifts are therefore accumulated on the reflection at both vacuum and substrate interfaces. In order to obtain these interfacial phase shifts, the Bohr-Sommerfeld quantization rule is applied to determine the energies of QWSs[19-21], offering a straightforward way to analyze the interfacial behavior and influences on the QWSs confined in the Pb nanostructures.

In this work, we have studied the growth and quantum well electronic states of the 2D Pb nanoislands on Nb-STO(001) by employing scanning tunneling microscopy and spectroscopy (STM/STS). According to dI/dU spectra taken on Nb-STO(001) substrate before and after Pb deposition, there is no significant changes observed on dI/dU spectra features, indicating the absence of the Pb wetting layer, such that the 2D Pb nanoislands are grown isolatedly on Nb-STO(001). At the low coverage of Pb, most of 2D Pb nanoislands are preferring an apparent height of 4 atomic layers because of a unique bilayer growth of Pb(111)[9-11]. On the other hand, these 2D Pb nanoislands start to merge into each other as visulaized directly by boundary lines on top of the large Pb isalnd at the higher coverage. From QWS energies as a function of Pb thickness, the band dispersion E(k) along the $\mathbf{\Gamma} - \mathbf{L}$ direction of bulk Pb(111) has been mapped out and the corresponding constant group velocity $v_g$ = 1.804 × 10$^6$ m/s and Fermi wavevector $k_F$ = 1.575 Å$^{-1}$ have bee extracted, respectively. In combination with phase accumulation model (PAM)[19-21], we are able to deduce the energy dependence of scattering phase shift $\varphi(E)$, which reveals the scattering interfaces of the Nb-STO(001) and the metallic Ag(111) are much alike.

**Experimental methods and calculation details:**

The experiments were performed in an ultrahigh vacuum (UHV) multifunctional chamber with the base pressure $p \leq 2 \times 10^{-10}$ mbar. The clean Nb-STO(001) substrate was prepared by cycles of degassing and flashing up to 1200 K through direct resistive heating. The Pb was evaporated on Nb-STO(001) at room temperature (RT) and the Pb deposition rate of 0.59 ML (ML, 1 ML = 0.286 nm = $d$, i.e., Pb(111) interlayer distance) per minute was calibrated from 2D Pb(111) nanoislands grown on Nb-STO(001). After preparation, the sample was transferred to a low-temperature scanning tunneling microscopy (LT-STM, Unisoku USM 1400S) held at 77 K for all measurements. The topography images were obtained from the constant-current mode with the bias voltage $U$ applied to the sample. For scanning tunneling spectroscopy (STS) measurements, a small bias voltage modulation $U_{mod}$ = 20 − 50 mV was added to U (frequency $v$ = 3991 Hz), such that tunneling differential conductance dI/dU spectra can be acquired by detecting the first harmonic signal by means of a lock-in amplifier. While measuring the dI/dU spectra at the constant current mode of spectroscopy, the feedback loop was off and the bias voltage was ramped at a fixed tunneling current value. On the contrary, the feedback loop was on and the bias voltage was ramped at a fixed tip-sample distance at the constant height mode of measurements.

The DFT calculations[22,23] were carried out within the generalized gradient approximation[24] with projector augmented wave (PAW) pseudopotential method[25], as implemented in the Vienna Ab-initio Simulation Package (VASP)[26,27]. The kinetic energy cutoff was set to 350 (25.7 Ry) and 500 eV (36.7 Ry) for Pb film and STO bulk, respectively. The Γ-centered 12 × 12 × 1 and 12 × 12 × 12 Monkhorst-Pack grid was used to sample the Brillouin zones for Pb film and STO bulk. Moreover, for all our surface calculations, the theoretical Pb and STO bulk lattice constant of 3.550 and 3.940 Å was adopted, respectively. We employed a periodically repeating slab consisting of 4 to 21 atomic Pb layers and a vacuum space of ∼20 Å. All Pb atoms were relaxed until the residual force was smaller than 0.001 eV/Å. Spin-orbit coupling was included in band structure and density of states calculations which is calculated by the integration near the Γ point.

**Results and discussions:**

The overview STM topography of about 1.8 ML Pb deposited on Nb-STO(001) at RT has been shown in Fig. 1(a), the hexagonal Pb nanoislands with quite uniform size and height have been grown and they are distributed homogenously over the whole surface of Nb-STO(001) at this coverage. Note that white arrows indicate where different step edges of Nb-STO(001) are. From the zoom-in image shown in Fig. 1(b), i.e., square frame marked by white dashed line in Fig. 1(a), the Pb nanoislands have atomically flat surfaces on top, and a line profile measured across one Pb nanoisland (blue dashed line from A to B) has been shown in Fig. 1(c). According to extracted apparent height in Fig. 1(c), the 1.11 nm corresponds to about 3.9 times 0.286 nm of Pb(111) interlayer distance, resulting in a thickness close to 4 ML. Since the Fermi wavelength $\lambda_F$ divided by the interlayer distance $d$ results in a unique ratio of bilayer periodicity ($\lambda_F/d \approx 4$) in the Pb(111)[9-11], the electron standing waves are thus easily to build up in these 2D Pb nanoislands and could manifest themselves into the formation of discrete QWSs.

With high spatial and energy resolutions, the STS offer an ideal tool to access the QWSs in both occupied and unoccupied energy regions. Distinct resonance peaks in tunneling conductance dI/dU spectra as a result of the electronic density of states (DOS) modulated by the QWSs can therefore be resolved. For example, the dI/dU curve acquired on 4ML Pb nanoislands (blue line) has been shown in Fig. 1(d), three peak features can be clearly identified at 1.0 eV, 3.2 eV and 4.2 eV, respectively. Besides QWSs on the 4ML Pb nanoislands, we have also measured dI/dU curves on Nb-STO(001) substrate before (red line) and after (black line) the deposition of Pb, in order to check whether there is a Pb wetting layer or not. As shown in Fig. 1(d), two peaks at 1.9 eV and 4.0 eV appear in both red and black lines, and they share the similar shape, which implies the absence of Pb wetting layer grown on Nb-STO(001) surface. To further support our findings, we conducted the DFT calculations on the band structure of the bulk STO, as shown in Fig. 1(e), to confirm the dI/dU curve measured on pristine Nb-STO(001) substrate without the deposition of Pb. From this band structure calculation, it is evident that two bands in the unfilled region (i.e., above Fermi level) with its local minima positioned at 1.9 eV and 4.0 eV show up along the $\Gamma$ point. In Fig. 1(f), we further calculate the density of states of bulk STO substrate by integrating the contribution of bands near the $\Gamma$ point with an energy broadening parameter $\Delta E$ = 0.1 eV. The same 1.9 eV and 4.0 eV peaks corresponding to the bands minima also appear in the calculated DOS and confirm those two peak features on the dI/dU curve (black line) measured from the bare Nb-STO(001) substrate. The consistency between the experimental dI/dU spectra (black line) and the DOS by DFT calculations (Fig. 1(f)) supports the absence of Pb wetting layer, meaning that the individual and separate 2D Pb nanoislands can be grown on Nb-STO(001).

To have an overall picture of growth behavior, we have studied the thickness-dependent sizes and heights of the 2D Pb nanoislands at various coverages of Pb deposited on Nb-STO(001) at RT. From Fig. 2(a) to (d), the STM overviews display how the growth behavior of 2D Pb nanoislands evolves with an increase of four representative Pb coverages. First of all, we can clearly see the 2D Pb nanoislands with hexagonal shape and a flattop are continuously to grow, and both the size and height keep getting increased up to 14.2 ML of Pb coverages in our results. Second, a single 2D Pb nanoisland always has a uniform thickness on top of the same substrate terrace at the RT growth, even when it's lateral size becomes a hundred nanometer wide, which is very different from either the crater structure found in Pb/Si(111)[28,29], or multilayer heights in Pb/Ag(111)[11] and Pb/Cu(111)[30] at various growth temperatures or even at the RT. From a closer look at the top surfaces of two large islands in Fig. 2(c) through enhancing the contrast of topography image, we have found a single large polygon Pb island can be formed by merging smaller hexagonal Pb islands into each other, whereas the boundary lines are indicated by black arrows.

The quantitative distributions of area occupation percentage of 2D Pb nanoislands as a function of thickness have been shown in the bottom of Fig. 2(a) to (d). In addition to the higher percentage of area occupation shifts to the higher thickness of islands when the Pb coverage is increasing, we have noticed that small 2D Pb nanoislands with a thickness of 4 atomic layers always exist although the area occupation percentage is dropping. This interesting observation explains that all the 2D Pb nanoislands are initially developed from the thickness of 4 atomic layers, and then they are not only getting thicker, but also increasing the area occupation simultaneously when the deposition of Pb coverage is increasing. Eventually, the coalescence occurs when the islands with the same thickness meet each other, and then merge into a single large island, which also explains the boundary lines visualized in Fig. 2(c) as described in previous paragraph. With systematic studies of thickness-dependent area occupation percentage, different types of 2D Pb nanoislands observed in previous studies can be therefore expected at the higher Pb coverage regime[7,8]. Note that the islands lacking a certain of thickness, for examples, thickness below the 4, the 5, the 17 and the 19 MLs are either rarely to be found in the experiments or the forbidden heights that could be attributed to the quantized nesting vectors confined on the Fermi surface of the Pb(111)[30].

After understanding the RT growth behavior of 2D Pb nanoislands on Nb-STO(001), we have carried out the STS measurements to access the thickness-dependent QWSs in both occupied and unoccupied energy regions. Two different operation modes, including the constant-height (C. H.) and the constant-current (C. C.) modes in the STS measurements (see experimental methods for details) have been applied to the energy ranges of -1.0 to +1.0 eV as well as +0.5 eV to +5.0 eV. The systematic results of dI/dU spectra have been arranged in the Fig. 3(a) and (b) for the even and the odd numbers of island thickness ($N$), respectively. Several resonance peak features in the dI/dU spectra are originated from the QWSs confined in 2D Pb nanoislands and they appear at discrete bias voltages corresponding to multiple quantized energy levels[9-11]. Interestingly, a special peak located at the energy close to 0.88 eV above Fermi level $E_F$ appears at all even-ML thickness (fig. 3(a)) and it almost does not show any energy shift when the thickness varies, which refers to the so called commensurate state as reported prevously[12,31]. Since this commensurate state has a bilayer periodicity, we have the $Q$ = 2 state in the wavevector of $(S/Q)(\pi/d)$ and the $S/Q$ must be an irreducible rational number, and $S$ and $Q$ are intergers[12,31]. In order to find the possible Fermi wave vector $k_F$ of Pb(111) to develop such commensurate state in the reduced Brillouin zone, the value of $S$ = 3 has to be determined for fulfilling the criteria[9-11].

Given that the commensurate state determined by the constant dI/dU peak feature at all even-ML thickness of the 2D Pb nanoislands, we are able to further assign the quantum numbers ($n$) for all QWSs in the simple particle-in-a-box model. Following the infinite-well approximation, we have the $k_n = (n\pi/Nd)$ together with already known wavevector of the commensurate state, the quantum numbers ($n$) have been assigned accordingly on top of each dI/dU peak in Fig. 3(a) and (b), respectively. By taking the energy and the wavevector of QWSs into account in the phase accumulation model (PAM) derived from basic ingredients of elementary multiple-reflection and nearly-free-electron theories[19-21], the energy dependent scattering phase shift $\varphi$ can be quantitatively obtained, which provides further insights on the QWS electrons scattered at separate boundaries including vacuum barrier and crystal interface of the 2D Pb nanoislands on Nb-STO(001).

Fig. 4(a) shows the arrangement of the energies of QWSs as a function of island thickness, the black and the red dots represent the results measured from the C. C. and the C. H modes of the STS, respectively. According to the Bohr-Sommerfeld quantization rule[21,32]:

$$2k(E)Nd + \varphi_B + \varphi_C = 2n\pi \quad (1)$$

where $k(E)$ is the energy-dependent wavevector of electron propagating along the $\Gamma - L$ direction of the Pb(111), and $\varphi = \varphi_B + \varphi_C$ is the total phase shift that combines vacuum barrier phase shift $\varphi_B$ and crystal interface phase shift $\varphi_C$ of the 2D Pb nanoislands on Nb-STO(001). In order to extract the $k(E)$ dispersion, in principle, we can apply the eq. (1) to each pair of the QWSs under the assumption of the same energy $E$, and then we can derive:

$$k(E) = \frac{n_2 - n_1}{N_2 - N_1}\frac{\pi}{d} \quad (2)$$

where the subscripts 1 and 2 stand for the QWSs at the two different thickness. However, in practice, there is a limit of energy resolution in experimental STS measurements and an energy uncertainty of $|E_2 - E_1| \leq 0.1$ eV has been taken into account for determining the energies of the QWSs. The results of the $k(E)$ dispersion (black dots) has been summarized in Fig. 4(b), which shows the linear relation between $E$ and $k$ along the $\Gamma - L$ direction in the extended Brillouin zone of the bulk Pb(111) band. According to the linear fit, the constant group velocity $v_g$ = 1.804 × 10$^6$ m/s and the Fermi wave vector $k_F$ = 1.575 ± 0.1 Å$^{-1}$ have been obtained, which is not only compaprable to the results of Pb/Ag(111)[11] (extracted green dots and orange line ), but also in a good agreement with previous photoemission studies[33,34]. Although the energy resolution of our STS measurements is better than 0.1 eV, we have also checked the lower energy uncertainty $|E_2 - E_1| \leq 0.05$ eV, and the output of the Fermi wave vector varies only within the error bar.

Since the *k(E)* dispersion has been obtained, we can further deduce the *φ(E)* by using the eq. (1). There are two components included in the total phase shift of *φ* = *φ$_B$* + *φ$_C$*, and they can be interpreted as following[35,36]:

$$\varphi_B = \pi \left( \sqrt{\frac{3.4 \text{ eV}}{E_V - E}} - 1 \right) \quad (3)$$

$$\varphi_C = \text{Re}\left[ -\cos^{-1}\left( 2\frac{E - E_L}{E_U - E_L} - 1 \right) \right] + \varphi_0 \quad (4)$$

where $E_V$ is the vacuum level, $E_U$ = 0 eV and $E_L$ = -3.9 eV are obtained from the conduction-band minimum and valence-band maximum at the Γ point of the Nb-STO(001) according to the tight-binding calculations[37], and *φ$_0$* denotes the phase offset. The results of the *φ(E)* accompanied by $E_V$ = 4.778 eV and *φ$_0$* = 0.3326 rad from the best fit of the *k(E)* in eq. (1) has been shown in Fig. 4(c) (black dots and blue line). Interestingly, the *φ(E)* of the Pb/Nb-STO(001) overall exhibits a dispersion behavior analogous to the Pb/Ag(111)[11] (extracted gray dots and orange line) in the entire energy range of Fig. 4(c), suggesting a metallic-like interface without an impact of the directional band gap and the Schottky barrier observed on the semiconductor of the Pb/Si(111)[38]. Another interesting feature of the *φ(E)* dispersion of the Pb/Nb-STO(001) is that the divergence of the *φ$_B$* becomes less prominent as compared to the Pb/Ag(111) in the higher energy range, indicating the high lying QWSs are less affected by the image potential[10] due to a larger $E_V$ value, i.e., 4.778 eV, and hence lower *φ$_B$* found in Pb/Nb-STO(001).

**Conclusions:**

In summary, we represent the systematic studies of the growth and the QWSs of the 2D Pb nanoislands on Nb-STO(001). According to the identical spectra features of the dI/dU curves taken on Nb-STO(001) before and after the Pb deposition, there is an absence of the Pb wetting layer, leading to the 2D Pb nanoislands are isolated from each other. In addition, the 2D Pb nanoislands prefer to develop from an apparent height of 4 atomic layers, and merge into a single large polygon Pb isalnd only when they meet with the same thickness at the RT growth. The thickness-dependent QWSs have been investigated in both occupied and unoccupied enegy regions by using the local probe of the STS measurements on the 2D Pb nanoislands. Given that the discrete energy positions of the QWSs and island thickness determined directly from experiemtns, the energy-dependent wave vector *k(E)* along the Γ − L direction of the Pb(111) has been obtained by means of the Bohr-Sommerfeld quantization rule. From a linear fit, the constant group velocity $v_g$ = 1.804 × 10$^6$ m/s and the Fermi wave vector $k_F$ = 1.575 ± 0.1 Å$^{-1}$ have been extracted, which are consistent with the values reported from different experiemntal methods in previous works. Furthermore, the total phase shift *φ(E)* from the electron scatterings between interfaces confined in the 2D Pb nanoislands on Nb-STO(001) display similar dispersion behavior with the Pb grown on metallic Ag(111). Lacking of a direct contact between individual 2D Pb nanoislands provides these nanoscale superconductors decoupled in real-sapce, enableing to investigate the mutual interlay between intrinic quantum confinement phenomena and type-II superconductivity of the Pb/Nb-STO(001).

**Acknowledgements:**

GYC, BYL, LWC, DSL and PJH acknowledge financial support from the competitive research funding from National Tsing Hua University, Ministry of Science and Technology of Taiwan under Grants No. MOST-108-2636-M-007-002 and MOST-107-2112-M-007-001-MY3, and center for quantum technology from the featured areas research center program within the framework of the higher education sprout project by the Ministry of Education (MOE) in Taiwan. PJH acknowledges the Prof. Jan-Chi Yang from the department of physics of National Cheng Kung University for the supply of the Nb-STO(001). CHH and FCC acknowledge support from the National Center for Theoretical Sciences and the Ministry of Science and Technology of Taiwan under Grants No. MOST-107-2628-M-110-001-MY3. CHH and FCC are also grateful to the National Center for High-Performance Computing for computer time and facilities.

**References**

1. J. F. Ge, Z. L. Liu, C. Liu, C. L. Gao, D. Qian, Q. K. Xue, Y. Liu, J. F. Jia, Superconductivity above 100 K in single-layer FeSe films on doped SrTiO3, Nat. Mater. **14** (2015) 285-289.
2. C. Tang, et al., Superconductivity dichotomy in K-coated single and double unit cell FeSe films on SrTiO3, Phys. Rev. B **92** (2015) 180507(R).
3. F. Li, et al., Interface-enhanced high-temperature superconductivity in single-unit-cell FeTe1−xSex films on SrTiO3, Phys. Rev. B **91** (2015) 220503(R).
4. G. Zhou, et al., Interface induced high temperature superconductivity in single unit-cell FeSe on SrTiO3(110), Appl. Phys. Lett. **108** (2016) 202603.
5. H. Ding, Y.-F. Lv, K. Zhao, W.-L. Wang, L. Wang, C.-L. Song, X. Chen, X.-C. Ma, Q.-K. Xue, High-temperature superconductivity in single-unit-cell FeSe films on anatase TiO2(001), Phys. Rev. Lett. **117** (2016) 067001.
6. Z. Shao, et al., Scanning tunneling microscopic observation of enhanced superconductivity in epitaxial Sn islands grown on SrTiO3 substrate, Sci. Bull. **63** (2018) 1332-1337.
7. Y. Yuan, X. Wang, C. Song, L. Wang, K. He, X. Ma, H. Yao, W. Li, Q.-K. Xue, Observation of coulomb gap and enhanced superconducting gap in nano-sized Pb islands grown on SrTiO3, Chin. Phys. Lett. **37** (2020) 017402.
8. H. Sun, et al., Scanning tunneling microscopic evidence of interface enhanced high-Tc superconductivity in Pb islands grown on SrTiO3, arXiv:1811.09395v1 (2018).
9. R. Feng, E. H. Conrad, Wetting-layer transformation for Pb nanocrystals grown on Si(111), Appl. Phys. Lett. **85** (2004) 3866-3868.
10. M. C. Yang, C. L. Lin, W. B. Su, S. P. Lin, S. M. Lu, H. Y. Lin, C. S. Chang, W. K. Hsu, Tien T. Tsong, Phase contribution of image potential on empty quantum well states in Pb islands on the Cu(111) surface, Phys. Rev. Lett. **102** (2009) 196102.
11. M. Becker, R. Berndt, Scattering and lifetime broadening of quantum well states in Pb films on Ag(111), Phys. Rev. B **81** (2010) 205438.
12. R. C. Jaklevic, J. Lambe, Experimental study of quantum size effects in thin metal films by electron tunneling, Phys. Rev. B **12** (1975) 4146-4160.
13. Y. Guo, et al., Superconductivity modulated by quantum size effects, Science **306** (2004) 1915-1917.
14. D. Eom, S. Qin, M.-Y. Chou, C. K. Shih, Persistent superconductivity in ultrathin Pb films: a

scanning tunneling spectroscopy study, Phys. Rev. Lett. **96** (2006) 027005.
15. C. Brun, I.-P. Hong, F. Patthey, I. Y. Sklyadneva, R. Heid, P. M. Echenique, K. P. Bohnen, E. V. Chulkov, W.-D. Schneider, Reduction of the superconducting gap of ultrathin Pb islands grown on Si(111), Phys. Rev. Lett. **102** (2009) 207002.
16. M. Hupalo, M. C. Tringides, Correlation between height selection and electronic structure of the uniform height Pb/Si(111) islands, Phys. Rev. B **65** (2002) 115406.
17. M. H. Upton, C. M. Wei, M. Y. Chou, T. Miller and T.-C. Chiang, Thermal stability and electronic structure of atomically uniform Pb films on Si(111), Phys. Rev. Lett. **93** (2004) 026802.
18. X. Ma, P. Jiang, Y. Qi, J. Jia, Y. Yang, W. Duan, W.-X. Li, X. Bao, S. B. Zhang, Q.-K. Xue, Experimental observation of quantum oscillation of surface chemical reactivities, Proc. Natl. Acad. Sci. **104** (2007) 9204-9208.
19. P. M. Echenique, J. B. Pendry, The existence and detection of Rydberg states at surfaces, J. Phys. C: Solid State Phys. **11** (1978) 2065-2075.
20. N. V. Smith, Phase analysis of image states and surface states associated with nearly-free-electron band gaps, Phys. Rev. B **32** (1985) 3549-3555.
21. T.-C. Chiang, Photoemission studies of quantum well states in thin films, Surf. Sci. Rep. **39** (2000) 181-235.
22. P. Hohenberg, W. Kohn, Inhomogeneous electron gas, Phys. Rev. **136** (1964) B864-B871.
23. W. Kohn, L. J. Sham, Self-consistent equations including exchange and correlation effects, Phys. Rev. **140** (1965) A1133-A1138.
24. J. P. Perdew, K. Burke, M. Ernzerhof, Generalized gradient approximation made simple, Phys. Rev. Lett. **77** (1996) 3865-3868.
25. G. Kresse, D. Joubert, From ultrasoft pseudopotentials to the projector augmented-wave method, Phys. Rev. B **59** (1999) 1758-1775.
26. G. Kresse, J. Hafner, *Ab initio* molecular dynamics for liquid metals, Phys. Rev. B **47** (1993) 558-561(R).
27. G. Kresse, J. Furthmüller, Efficient iterative schemes for *ab initio* total-energy calculations using a plane-wave basis set, Phys. Rev. B **54** (1996) 11169.
28. W. B. Su, S. H. Chang, W. B. Jian, C. S. Chang, L. J. Chen, Tien T. Tsong, Correlation between quantized electronic states and oscillatory thickness relaxations of 2D Pb islands on Si(111)-(7×7) surfaces, Phys. Rev. Lett. **86** (2001) 5116-5119.
29. K. Wang, X. Zhang, M. M. T. Loy, T.-C. Chiang, X. Xiao, Pseudogap mediated by quantum-size effects in lead islands, Phys. Rev. Lett. **102** (2009) 076801.
30. R. Otero, A. L. Vázquez de Parga, R. Miranda, Observation of preferred heights in Pb nanoislands: A quantum size effect, Phys. Rev. B **66** (2002) 115401.
31. R. C. Jaklevic, J. Lambe, M. Mikkor, W. C. Vassell, Observation of electron standing waves in a crystalline box, Phys. Rev. Lett. **26** (1971) 88-92.
32. M. Milum, P. Pervan, D. P. Woodruff, Quantum well structures in thin metal films: simple model physics in reality?, Rep. Prog. Phys. **65** (2002) 99-141.
33. Y.-F. Zhang, J.-F. Jia, T.-Z. Han, Z. Tang, Q.-T. Shen, Y. Guo, Z. Q. Qiu, Q.-K. Xue, Band structure and oscillatory electron-phonon coupling of Pb thin films determined by atomic-layer-resolved quantum-well states, Phys. Rev. Lett. **95** (2005) 096802.
34. J. R. Anderson, A. V. Gold, Fermi surface, pseudopotential coefficients, and spin-orbit coupling in lead, Phys. Rev. **139** (1965) A1459-A1481.
35. E. G. McRae, M. L. Kane, Calculations on the effect of the surface potential barrier in LEED, Surf. Sci. **108** (1981) 435-445.

36. L. Rettig, P. K. Kirchmann, and U. Bovensiepen, Ultrafast dynamics of occupied quantum well states in Pb/Si(111), New J. Phys. **14** (2012) 023047.
37. M. Takizawa, K. Maekawa, H. Wadati, T. Yoshida, A. Fujimori, H. Kumigashira, M. Oshima, Angle-resolved photoemission study of Nb-doped SrTiO3, Phys. Rev. B **79** (2009) 113103.
38. S. Pan, Q. Liu, F. Ming, K. Wang, X. Xiao, Interface effects on the quantum well states of Pb thin films, J. Phys.: Condens. Matter. **23** (2011) 485001.

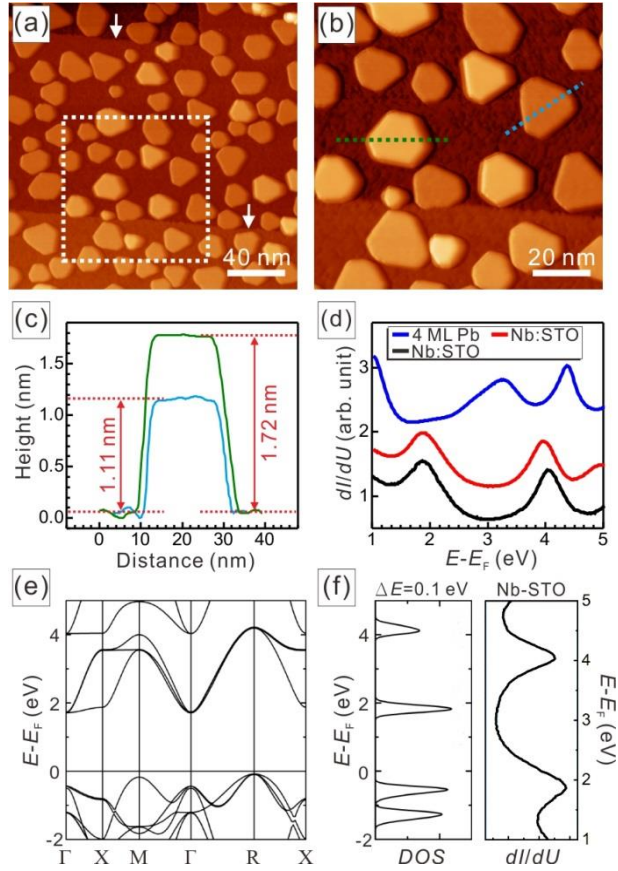

**Fig. 1.** STM topography and STS measurements of Pb/Nb-STO(001) (a) An overview image of 2D Pb nanoislands grown Nb-STO(001) with a Pb coverage of 1.77 ML. Two step edges of Nb-STO(001) substrate are indicated by white arrows. (b) The zoom-in image marked by the white dashed square in (a). Two kinds of apparent heights of 2D Pb nanoislands have been found at this Pb coverage, and their corresponding line profiles have been shown in (c). (c) The line profiles of 2D Pb nanoislands measured from (b), and their apparent heights are marked by the blue and green lines, respectively. The apparent heights of 1.11 nm and 1.72 nm refer to 4 ML and 6 ML interlayer distance of the Pb(111). (d) The dI/dU spectra taken on 2D Pb nanoislands (blue) with 4 ML apparent height and Nb-STO(001) substrate before (black) and after (red) Pb deposition. The spectra features, i.e., peaks at 1.9 eV and 4.0 eV, are identical in shape and bias voltages of Nb-STO(001) substrate before and after the Pb deposition, indicating the absence of Pb wetting layer formation. (e) Band structure for bulk STO and (f) its DOS calculated by the integration near the $\Gamma$ point, which is consistent with the experimental dI/dU peak features observed at 1.9 eV and 4.0 eV.

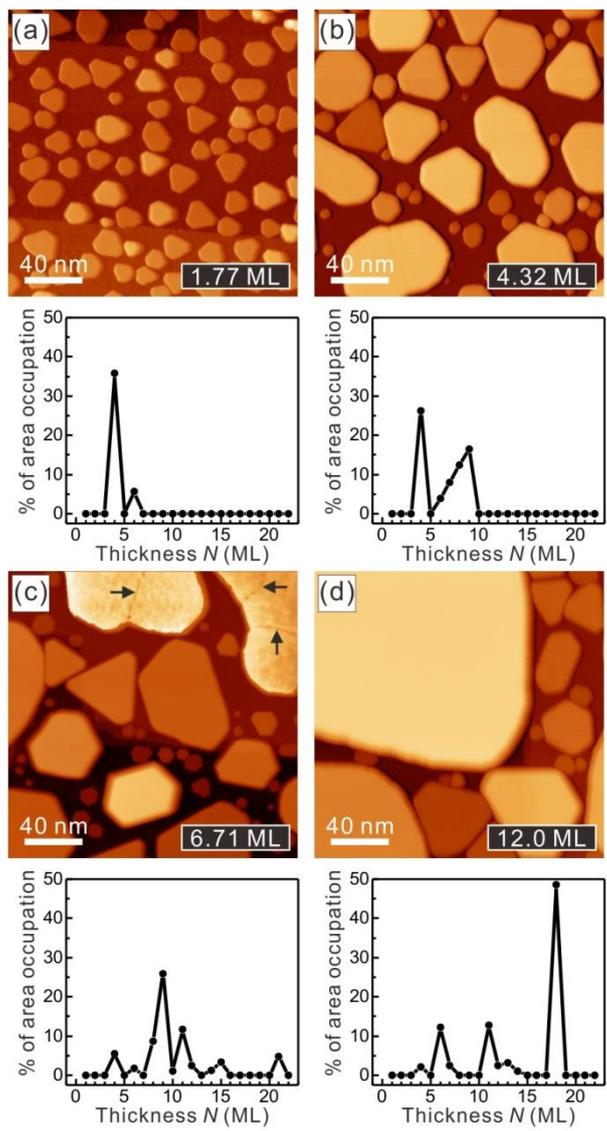

**Fig. 2.** Percentage of area occupation as a function of the Pb coverages. (a)-(d) STM overviews images at various Pb coverages. The area sizes as well as apparent height of the 2D Pb nanoislands are getting larger with an increase of Pb coverages. In addition, as shown in (c) by three black arrows, the coalescence of the 2D Pb nanoislands occurs only when they meet with the same apparent height, as the boundary lines visualized directly on top of a single large Pb island. According to the statistics of thickness-dependent area occupation percentage, the 2D Pb nanoislands always develop from the building blocks with an apparent height of 4 atomic layers.

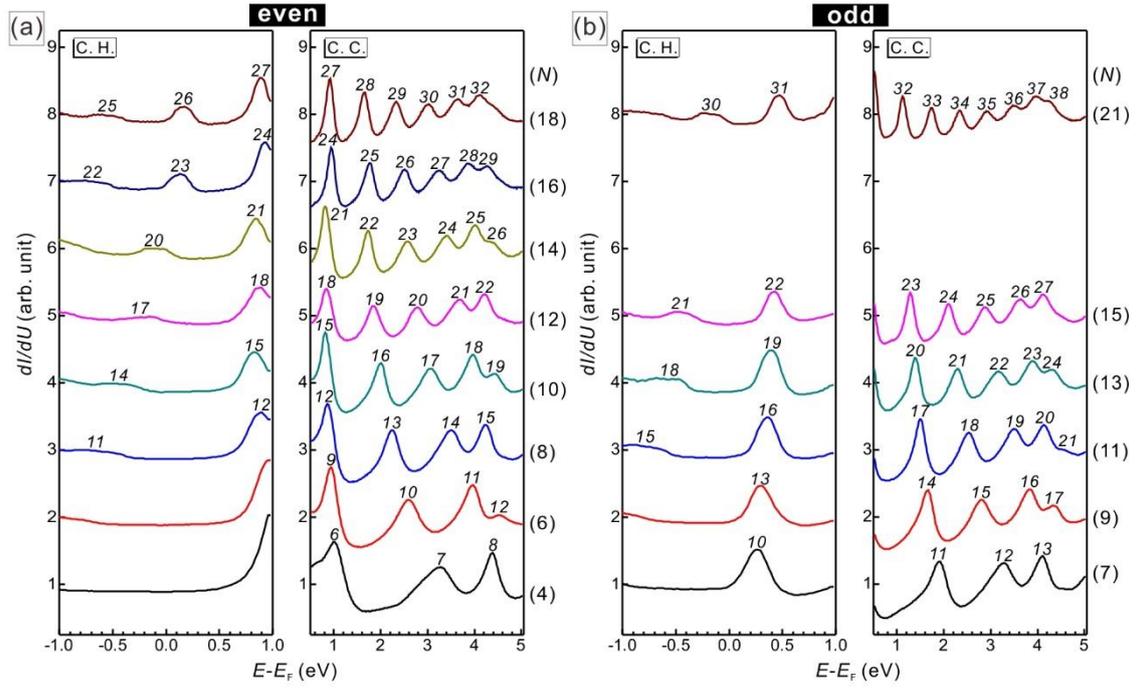

**Fig. 3.** Thickness-dependent QWSs measured on the 2D Pb nanoislands in both occupied and unoccupied energy regions. The constant-current (C.C) and constant-height (C. H.) dI/dU spectra have been arranged into (a) even- and (b) odd-ML of island thickness $N$. The quantum numbers $n$ have been determined from the commensurate state resolved at 0.88 eV of the 2D Pb nanoislands with the even-ML thickness.

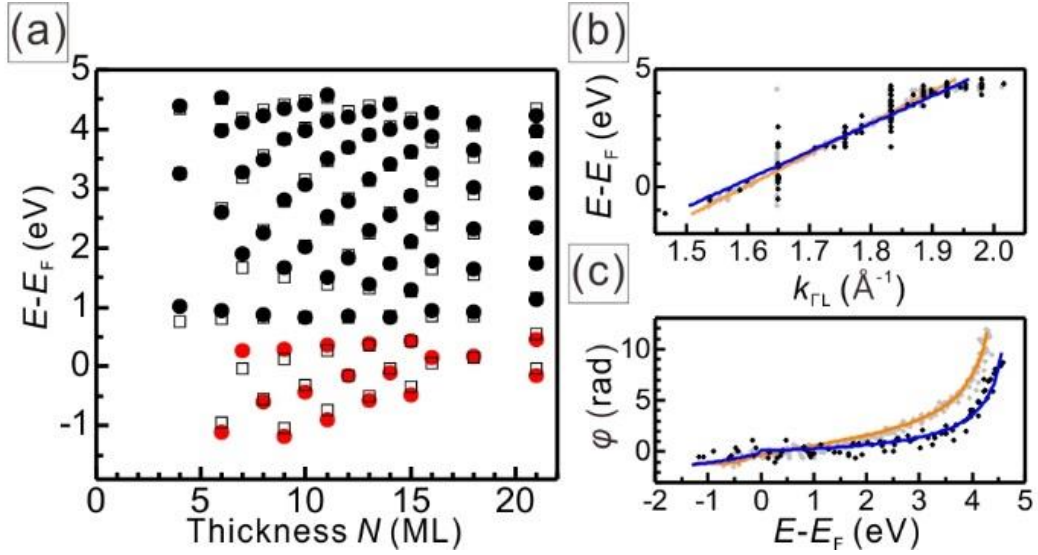

**Fig. 4.** The PAM on the QWSs of the 2D Pb nanoislands on Nb-STO(001). (a) QWSs energies as a function of island thickness measured from dI/dU spectra. Red and black dots represent the data measured from C. H. and C. C. modes of STS, respectively. The black empty squares are corresponding results calculated according to the PAM. (b) The $k(E)$ of the Pb/Nb-STO(001) has been obtained from applying Bohr-Sommerfeld quantization rule to experimentally extracted QWSs energies (black dots and blue line), showing a linear dispersion of the $k(E)$ along $\Gamma - L$ direction in the extended Brillouin zone of Pb(111). A linear fit gives the constant group velocity $v_g = 1.804 \times 10^6$ m/s and the Fermi wave vector $k_F = 1.575 \pm 0.1$ Å$^{-1}$, which are in line with data extracted from the Ref. Ref. 11 of the Pb/Ag(111) (gray dots and orange line). (c) Energy-dependent total phase shift $\varphi(E)$ (black dots and blue line) overall displays similar dispersion with the extracted data (gray dots and orange line) from the Ref. Ref. 11 of the Pb/Ag(111), but smaller divergence at the higher energy region due to the less impact on the high lying QWSs from the image potential.